\documentclass[lettersize,journal]{IEEEtran}
\usepackage{amsmath,amsfonts}
\usepackage{algorithmic}
\usepackage{array}
\usepackage[caption=false,font=normalsize,labelfont=sf,textfont=sf]{subfig}
\usepackage{textcomp}
\usepackage{stfloats}
\usepackage{url}
\usepackage{verbatim}
\usepackage{graphicx}
\def\BibTeX{{\rm B\kern-.05em{\sc i\kern-.025em b}\kern-.08em
    T\kern-.1667em\lower.7ex\hbox{E}\kern-.125emX}}
\usepackage{balance}
\usepackage{multirow}
\usepackage{booktabs}
\usepackage{xcolor}
\usepackage{caption}
\usepackage{hyperref}
\usepackage{listings}
\usepackage{etoolbox}
\makeatletter
\patchcmd{\@makefntext}{\hss\@makefnmark}{\@makefnmark\ }{}{}
\makeatother

\newcommand{\myparagraph}[1]{{\vspace{.2em} \noindent\hspace{0.5em}\bf #1}}

\begin{document}
\title{Exploring the Rate-Distortion-Complexity Optimization in Neural Image Compression}
\author{Yixin Gao, Runsen Feng, Zongyu Guo, Zhibo Chen, \it{Senior Member, IEEE}
\thanks{
%*Yixin Gao and Runsen Feng contribute equally to this work.

Yixin Gao, Runsen Feng, Zongyu Guo and Zhibo Chen are with the Department of Electronic Engineer and Information Science, University of Science and Technology of China, Hefei, Anhui, 230026, China. Corresponding Author: Zhibo Chen (chenzhibo@ustc.edu.cn).}}

\markboth{Journal of \LaTeX\ Class Files,~Vol.~18, No.~9, September~2020}%
{How to Use the IEEEtran \LaTeX \ Templates} 

\maketitle

\begin{abstract}
Despite a short history, neural image codecs have been shown to surpass classical image codecs in terms of rate-distortion performance. However, most of them suffer from significantly longer decoding times, which hinders the practical applications of neural image codecs. This issue is especially pronounced when employing an effective yet time-consuming autoregressive context model since it would increase entropy decoding time by orders of magnitude. In this paper, unlike most previous works that pursue optimal RD performance while temporally overlooking the coding complexity, we make a systematical investigation on the rate-distortion-complexity (RDC) optimization in neural image compression. By quantifying the decoding complexity as a factor in the optimization goal, we are now able to precisely control the RDC trade-off and then demonstrate how the rate-distortion performance of neural image codecs could adapt to various complexity demands. Going beyond the investigation of RDC optimization, a variable-complexity neural codec is designed to leverage the spatial dependencies adaptively according to industrial demands, which supports fine-grained complexity adjustment by balancing the RDC tradeoff. By implementing this scheme in a powerful base model, we demonstrate the feasibility and flexibility of RDC optimization for neural image codecs.
\end{abstract}


\section{Introduction}
\IEEEPARstart{T}{ime} complexity is a critical factor in the practical application of lossy image codecs. Nowadays, the widely used classical image codecs like HEVC/H.265~\cite{sullivan2012overview} intra and VVC/H.266~\cite{ohm2018versatile} intra, achieve great rate-distortion performance that relies on manually designed complex coding modes. Therefore, the encoding process for these traditional image codecs would become exceedingly time-consuming, as they usually search among a large group of coding modes for better performance.
To realize the practical applications of the image codecs, numerous studies have been conducted to reduce the encoding complexity~\cite{shen2014effective,shi2019asymmetric,dong2021fast,zhu2017binary,kuanar2019adaptive,wu2022hg}, which may sacrifice rate-distortion performance. In fact, the rate-distortion-complexity performance of a codec is more important in practice, and it is desired to control the RDC tradeoff flexibly.
\begin{figure}[htbp]
  \centering
  \includegraphics[width=1.0\linewidth]{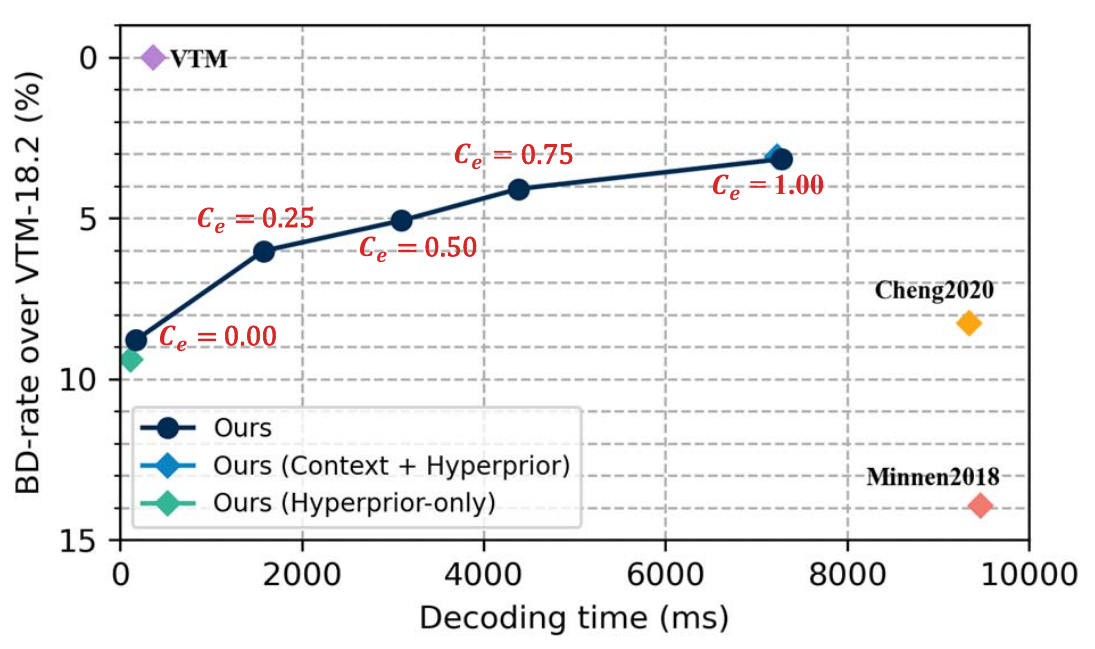}
  \caption{BD-rate (smaller is better) vs decoding time on Kodak.  
  $C_e=0$ corresponds to fast decoding, while $C_e=1$ is slowly decoding. Our variable-complexity model could finely adjust decoding time, controlled by a required value of $C_e$.} 
  \label{fig:BD_rate_vs_DecodingTime}
\end{figure}

When it comes to the field of neural image compression, the primary obstacle hindering widespread adoption predominantly lies in decoding~\cite{pan2021analyzing}. 
Contrary to classical image compression, these neural image codecs can encode images very fast especially when GPU acceleration is employed. Over the past few years, we have witnessed a rapid development of neural image compression methods. Some of them~\cite{balle2017end,minnen2018joint,cheng2020learned,guo2021soft} have achieved better rate-distortion performance than classical image compression standards including JPEG2000~\cite{rabbani2002overview}, BPG~\cite{bellard2015bpg}, and even the state-of-the-art (SOTA) classical codec VVC intra. 
A bunch of previous works demonstrates that employing a complex yet effective entropy model is critical to achieving high rate-distortion performance, albeit at the cost of unacceptably long decoding time~\cite{minnen2018joint,Lee2019Context,cheng2020learned,zhou2019multi,qian2020learning,guo2021causal}.
For instance, the popular context model-based neural image codec \cite{minnen2018joint,Lee2019Context} would bring approximately 7\% rate savings, but require more than 50$\times$ decoding time for a 768$\times$512 image due to the autoregressive nature of this entropy model as shown in Fig.~\ref{fig:BD_rate_vs_DecodingTime} (compare the Context + Hyperprior model with the Hyperprior-only model). To facilitate practical applications, some recent studies~\cite{li2020efficient,minnen2020channel,he2021checkerboard,chen2021end,he2022elic} have explored some effective alternatives to the spatial autoregressive entropy model. 
Different from solely pursuing high rate-distortion (RD) performance, people now start to pay more attention to developing a practical neural image codec that keeps a balance between rate, distortion, and complexity.

In this paper, we conduct a systematic investigation of the optimization of rate-distortion-complexity (RDC) for neural image compression. Considering that a majority of the coding time is consumed by entropy decoding in those prevalent context-model-based codecs, we first try to quantify the decoding complexity by counting the sequential coding frequency of the autoregressive model with a binary mask. This mask determines whether a spatial position is entropy-coded with sequential prediction or not. 
As a result, the sum value of this binary mask reveals the times that we use the context model, therefore is directly correlated to the decoding complexity.
Although mask generation can be manually designed or end-to-end optimized, our key observation is that mask convolution \cite{van2016pixel} is primarily required at object edges and in areas with complex contexts.
To achieve better RD performance under a complexity constraint, we incorporate decoding complexity as a term in the optimization objective, allowing us to precisely control the RDC trade-off.
We then show how RD curves could be finely adjusted to accommodate various demands of decoding complexity, from rapid to arbitrarily slower decoding.
To the best of our knowledge, this is the first time that RDC optimization is explored in neural image compression.

Going beyond the investigation of RDC optimization, we propose a unified compression model capable of supporting finely specified complexity requirements through a two-stage optimization process. 
In the first stage, inspired by the significant success of masked image modeling, such as BeiT~\cite{bao2021beit}, MAE~\cite{he2022masked}, and SimMIM~\cite{xie2022simmim}, we try to build a pre-trained model that supports arbitrary spatial mask input. We employ conditional transform modules that take the user-desired complexity indicator $C_e$ as a condition to modulate the input feature. In the second stage, to pursue better RDC tradeoffs, the adaptive spatial mask under each given complexity factor is learned by jointly optimizing RDC constraints.
The mask is decoded from the side information at the decoder side, eliminating the need for an explicit overhead transmission.
Consequently, our variable-complexity codec can adaptively control the spatial dependencies modeling in the context model.
As illustrated in Fig.~\ref{fig:BD_rate_vs_DecodingTime}, our model enables fine-grained adjustment of decoding complexity.
By implementing rate-distortion-complexity optimization in a powerful base model, which could be further integrated with prevailing neural image codecs, we demonstrate its considerable potential for both research and industrial applications.

In short, our contributions can be summarized as follows: 
\begin{enumerate}
\item {By quantifying the complexity of entropy decoding into optimization, for the first time, we are able to finely control the rate-distortion-complexity trade-off for neural image codecs.}

\item {We introduce a unified variable-complexity image compression model that can adjust the decoding complexity to a fine granularity within a single model.} 

\item {We implement the variable-complexity scheme in a powerful base model and conduct comprehensive experiments that verify the potential of RDC optimization in neural image compression.} 
\end{enumerate}

The rest of the paper is organized as follows. Section~\ref{sec:RelatedWork} reviews related literature on traditional image compression, deep image compression, and complexity-constrained rate-distortion optimization. 
Section~\ref{sec:RDCOoptimization} describes the definition of quantified decoding complexity in the entropy model and the rate-distortion-complexity optimization.
Section~\ref{sec:VCModel} presents the proposed method in detail, while experimental results and analysis are shown in Section~\ref{sec:Experiments}.

\section{Related Work}
\label{sec:RelatedWork}
\subsection{Traditional Image Compression}
Traditional image compression standards, such as JPEG~\cite{wallace1992jpeg}, JPEG2000~\cite{rabbani2002overview}, HEVC Intra~\cite{sullivan2012overview}, and VVC Intra~\cite{ohm2018versatile}, have been extensively used in practice after several decades of development. The current mainstream codecs, HEVC and VVC, are both based on block partitioning, which significantly impacts the rate-distortion (RD) performance improvement and complexity increase. On one hand, the partition search (PS) for optimal partition structures is time-consuming, as it goes through most of the encoding process for each coding unit. On the other hand, more flexible block partitioning structures and richer intra-mode decisions  lead to higher coding efficiency and computational complexity.

In recent years, complexity-constrained RD optimization methods have been widely considered to facilitate practical applications, falling into two categories: PS-dependent~\cite{shen2014effective,min2014fast,shi2019asymmetric,yang2019low,tang2019adaptive,cui2020gradient,saldanha2021learning,dong2021fast} and PS-free approaches~\cite{liu2016cu,zhu2017binary,kuanar2019adaptive,wu2022hg}.
PS-dependent approaches focus on mode skipping or early termination, while PS-free approaches aim to predict the full partition structure without the PS process of the encoder. 
Furthermore, some works~\cite{cho2013fast,zhang2018complexity,li2021deepqtmt,chen2020learned,huang2021modeling,saldanha2021configurable,feng2023partition,hosseini2021fine} control complexity for various industry requirements. Among them, \cite{hosseini2021fine} proposes fine-grained complexity control for numerous video-capable devices that operate on batteries with limited and varying processing power.

\subsection{Learned Image Compression}
Over the past few years, learned image compression methods \cite{he2022elic,guo2021causal,wu2021learned,cheng2020learned,minnen2020channel,minnen2018joint,balle2018variational} based on nonlinear transform coding \cite{balle2020nonlinear} have achieved comparable or even better performance than traditional coding standards like HEVC \cite{sullivan2012overview} and VVC \cite{ohm2018versatile}, which have been developed for decades. However, many of them focus on improving rate-distortion performance by designing more computationally expensive autoregressive (AR) entropy models, resulting in a significant increase in decoding time. To reduce decoding complexity while retaining RD performance, some previous works \cite{he2021checkerboard,minnen2020channel} propose computationally efficient AR models by regrouping the reference context in checkerboard \cite{he2021checkerboard} or channel-wise autoregressive \cite{minnen2020channel} manners, substantially reducing the corresponding AR complexity. While these methods achieve a good trade-off between rate-distortion and complexity, the RDC optimization is not systematically considered, and moreover, their decoding complexity is not adjustable for different application scenarios. 

\section{The Rate-distortion-complexity Optimization}
\label{sec:RDCOoptimization}
\subsection{Learned Image Compression Framework}
The objective of lossy image compression is to minimize the distortion $D$ between the reconstructed and original image as much as possible under the constraints of limited transmission rate $R$, which can be formulated by the Lagrange multiplier method:
\begin{equation}
\boldsymbol {\mathcal{L}} = R + \lambda \cdot D.
\label{eq:RD}
\end{equation}

For neural image compression, prevailing methods are based on variational autoencoders with an entropy bottleneck \cite{balle2016end}. In this framework, the encoder first converts an origin image $\boldsymbol x$ into a latent representation $\boldsymbol y$, which is then quantized to discrete representation $\boldsymbol {\hat{y}}$. To losslessly compress {$\boldsymbol {\hat{y}}$} into a bitstream, an entropy model will estimate the probability distribution of the quantized latent representation $\boldsymbol {\hat{y}}$ {\textit{i.e.}, $p_{\boldsymbol {\hat{y}}}$}. The entropy model is shared at both the encoder and decoder sides. % Since it dominates the bitrate, an effective entropy model is vitally important.
After transmission, at the decoder side, the discrete latent representation $\boldsymbol {\hat{y}}$ is mapped to pixels $\boldsymbol {\hat{x}}$. The training goal in Eq.\ref{eq:RD} can be further specified in a VAE style:
\begin{equation}
\boldsymbol {\mathcal{L}} = \mathbb{E}_{\boldsymbol x \sim p_{\boldsymbol x} } \left [  -\log_{2}{p_{\boldsymbol {\hat{y}} }\left ( \boldsymbol {\hat{y}} \right ) } \right ] + \lambda \cdot \mathbb{E}_{\boldsymbol x \sim p_{\boldsymbol x} } \left [ d\left ( \boldsymbol x, \boldsymbol {\hat{x}} \right )  \right ],
\label{eq:BasicLoss1}
\end{equation}
where $d\left ( \boldsymbol x, \boldsymbol {\hat{x}} \right )$ represents the distortion value under a given metric, such as mean squared error (MSE). 
The work of \cite{balle2018variational} introduces a hyperprior entropy model, which leverages side information $\boldsymbol z$ to capture spatial structure information of the latent representation $\boldsymbol y$, formulated by
\begin{align}
\boldsymbol {z} &= h_{a}(\boldsymbol {y};\boldsymbol{\phi}_h), \nonumber \\
\boldsymbol {\hat{z}} &= Q(\boldsymbol {z}), \\
p_{\boldsymbol {\hat{y}} \mid \boldsymbol {\hat{z}} }\left ( \boldsymbol {\hat{y}} | \boldsymbol {\hat{z}} \right ) &\gets h_{s}(\boldsymbol {\hat{z}};\boldsymbol{\theta}_h). \nonumber
% \label{eq:HyperLoss}
\end{align}
where $h_a$ and $h_s$ denote the hyper encoder and hyper decoder, and $\boldsymbol{\phi}_h$ and $\boldsymbol{\theta}_h$ are the network parameters. The distribution of $\boldsymbol {\hat{y}}$ could be parameterized conditioned on $\boldsymbol {\hat{z}}$, \textit{e.g.}, a conditional Mean \& Scale Gaussian distribution \cite{minnen2018joint}.
The discrete entropy of $\boldsymbol {\hat{z}}$ is estimated by an unconditional factorized entropy model $p_{\boldsymbol {\hat{z}} } $.

Furthermore, to better reduce the spatial dependencies in neighboring latent representation, 
%While the hyperprior model can partially remove spatial redundancies in $\boldsymbol {\hat{y}}$, there remain local dependencies in neighboring latent representation. To better reduce these spatial redundancies, 
the works of \cite{minnen2018joint,Lee2019Context} propose using an autoregressive context model $g_{cm}$ 
to predict the unknown latent based on the already decoded values: 
\begin{align}
{\boldsymbol {\hat{y}}}_{<{i}} &= \left \{ {\boldsymbol {\hat{y}}}_{1},\dots, {\boldsymbol {\hat{y}}}_{i-1} \right \} \nonumber \\
\boldsymbol {\Phi}_{cm,i}&=g_{cm}({\boldsymbol {\hat{y}}}_{<{i}}; \boldsymbol {\theta}_{cm}),
\label{eq:MaskConvLoss}
\end{align}
where ${\boldsymbol {\hat{y}}}_{<{i}}$ represents the causal context of $\boldsymbol {\hat{y}}_i$. Each context representation $\boldsymbol {\Phi}_{cm,i}$ is used to jointly predict entropy parameters in conjunction with the hyperprior $\boldsymbol {\hat{z}}$.
For practical applications, the autoregressive context model, which relies on mask convolution \cite{van2016conditional}, is an unfavorable choice for decoding, as it requires computing convolution serially $H\times W$ times when decoding a $C\times H\times W $ feature map. In contrast, although the pure hyperprior model is inferior to the joint hyperprior and context model~\cite{minnen2018joint}, it can be implemented in a parallel coding process. In short, incorporating the context model brings an obvious gain in RD performance but increases the decoding time by orders of magnitude at the same time.

% \subsection{Quantifying the Complexity of Entropy Decoding}
\subsection{Description of Decoding Complexity}
\label{sec:EntropyModelComplexity}
The concerns in practical applications of an image codec are not merely the rate-distortion (RD) performance. Coding complexity, especially decoding complexity, affects the user experience in many real-time applications. Consequently, it is crucial to take into account the decoding complexity alongside the RD optimization. We extend the RD optimization {(Eq.~\ref{eq:RD})} to the following rate-distortion-complexity function, aiming to minimize RD loss under a decoding time constraint $T_{t}$:
\begin{equation}
\min R+\lambda \cdot D \quad s.t.\quad  T\le T_{t},
\label{eq:RDC}
\end{equation}
where $T$ denotes the decoding time. For neural image codecs, we can generally split the decoding time into two components:
\begin{equation}
T= T_{e} + T_{n}.
\label{eq:T2Teo}
\end{equation}
Here $T_{e}$ denotes the time of entropy decoding and $T_{n}$ represents the time of parallel network inference for decoding, which includes the upsampling transform in the entropy model and the decoder transform. It is known that usually $T_{e}$ is larger than $T_{n}$, especially when using an autoregressive context model for entropy coding \cite{pan2021analyzing}, as latent representation decoding must be done serially due to the autoregressive nature of mask convolution~\cite{van2016conditional}. 
%For the image compression model without an autoregressive context model~\cite{balle2018variational}, $T_{e} > T_{n}$ as the network inference can be computed rapidly on GPUs, while entropy decoding should be operated on CPUs~\cite{pan2021analyzing}. As a result, the time spent on entropy decoding, \textit{i.e.}, $T_{e}$, would increase by orders of magnitude. 
%\gyx{It is known that usually $T_{e}$ is larger than $T_{n}$ no matter whether there is an autoregressive context model for entropy coding \cite{pan2021analyzing}. For the image compression model \cite{balle2018variational} without an autoregressive context model, $T_{e} > T_{n}$ because the network inference can be computed rapidly on GPUs, while entropy decoding should be operated on CPUs~\cite{pan2021analyzing}.}
%with frequent data transmission between CPUs and GPUs \cite{pan2021analyzing}. 
%While for the image compression model containing a vanilla context model \cite{minnen2018joint,Lee2019Context}, the situation worsens, as latent representation decoding must be done serially due to the autoregressive nature of mask convolution~\cite{van2016conditional}.
To develop a faster entropy model, some works~\cite{he2021checkerboard,minnen2020channel,he2022elic} explore changing dependencies modeling in latent representation. Regardless of the architecture of the entropy model, we can describe the complexity of entropy decoding according to the number of autoregressive times $n$, which is O(n).
Specifically, the initial hyperprior model~\cite{balle2018variational} has $n=1$. The number of  autoregressive times in checkerboard context model~\cite{he2021checkerboard} is 2. In the channel-wise autoregressive model~\cite{minnen2020channel}, the number of autoregressive times is $s$, where $s$ is the number of channel groups. Following this definition, for the first time, it is possible to quantitatively describe the decoding complexity of an entropy model.

% Some prior works~\cite{he2021checkerboard,minnen2020channel,he2022elic} explore changing dependencies modeling in latent representation to develop a faster entropy model. 
% No matter which architecture the entropy model has, 

\subsection{The Proposed RDC Optimization Method}
After analyzing the complexity of entropy decoding, we can formulate a novel optimization goal to balance the rate-distortion-complexity (RDC) tradeoff. Since the decoding time spent for network inference, $T_n$, is relatively negligible and directly correlated with the computing speed of hardware, we omit this component in the RDC loss function. The total RDC optimization goal is:
\begin{equation}
\mathcal{L}=R+{\lambda}_{D} \cdot D + {\lambda}_{C} \cdot C_{e},
\label{eq:RDC_goal}
\end{equation}
where $C_{e}$ is an indicator of entropy decoding time and another Lagrange multiplier ${\lambda}_{C}$ is imposed to control the time complexity of entropy decoding. In Section \ref{sec:EntropyModelComplexity}, we mentioned that the time complexity of entropy decoding is strongly correlated to the autoregressive nature of the entropy model. Here, based on the typical spatial autoregressive context model \cite{minnen2018joint,Lee2019Context}, we can further compute the value of $C_{e}$, as introduced below.
 
An important intuition is that not all the spatial positions must be predicted with autoregressive mask convolution. It is true that a more accurate estimation of the entropy of the latent representation is achieved with the help of mask convolution. However, in some cases such as decoding a smooth image full of sky, the hyperprior variable ${\boldsymbol z}$ is already sufficient to accurately estimate distribution parameters. In other words, we can sometimes avoid using the time-consuming mask convolution, depending on the specific image context. In short, we can generate an additional mask to indicate which spatial positions should be predicted serially. 
We denote this mask as $\boldsymbol M$, where ${\boldsymbol M}_{j}=1$ represents that the distribution probability of latent ${\boldsymbol {\hat y}}_{j}$ is estimated with the help of mask convolution. Conversely, ${\boldsymbol M}_{j}=0$ means that ${\boldsymbol {\hat y}}_{j}$ is decoded in parallel together with other latent representation that has the same zero mask value. Since more positions using mask convolution lead to longer decoding time, we can factorize the indicator of entropy decoding complexity $C_{e}$ as
\begin{equation}
%C_{e} = \frac{\sum_{i=0}^{H-1} \sum_{j=0}^{W-1} {\boldsymbol M}_{i,j}}{H \times W}.
C_{e} = \frac{\sum_{j=0}^{H \times W -1} {\boldsymbol M}_{j}}{H \times W},
\label{eq:T2Teo}
\end{equation}
where $H,W$ are the height and width of latent ${\boldsymbol {\hat y}}$, proportional to the original image resolution. An all-zero mask is equivalent to a pure parallel hyperprior entropy model \cite{balle2018variational}, while an all-one mask is equivalent to a pure mask convolution-based context model \cite{minnen2018joint,Lee2019Context}.

To generate this spatial mask ${\boldsymbol M}$, both manually designing or end-to-end optimization are feasible solutions. However, the handcrafted masks relatively lack adaptability to image content, which may degrade the RD performance. Therefore, we incorporate decoding complexity as a term in the optimization objective to learn the mask for better adaptability and performance.
To obtain this spatial mask ${\boldsymbol M}$ at the decoder side, we can decode the mask from the hyperprior ${\boldsymbol z}$. Usually, a small number of bits are spent to transmit the side information ${\boldsymbol z}$, which helps us transmit the spatial mask with negligible extra bits. Since the spatial mask is binary, we adopt the Gumbel-softmax reparameterization trick~\cite{jang2016categorical} to learn this binary mask and perform backpropagation. By generating such a learnable spatial mask, the entropy model can adaptively control the decoding complexity through end-to-end optimization. 
Combing Eq. \ref{eq:RDC_goal} and Eq. \ref{eq:T2Teo}, we can easily deduce that if we impose a larger constraint on the term $C_e$, \textit{i.e.}, larger $\lambda_{C}$ value in the optimization goal, there will be more positions masked with zero-value. Consequently, the time complexity of entropy decoding would be smaller.

Now we are able to finely control the RDC tradeoff in neural image compression by adjusting the hyperparameter $\lambda_{C}$ in the optimization goal. It is the first time, to our knowledge, that RDC optimization is studied for neural image compression.

% \vspace{0.8cm}
\section{Variable-complexity Image Compression}
\label{sec:VCModel}
\subsection{Overview}
% Overview
Based on our introduction in Section \ref{sec:RDCOoptimization}, we can train compression models under specific constraints of decoding complexity. However, for practical usage, it is of great importance to control the decoding complexity after acknowledging the latency tolerance at the decoder side. Since we do not expect to store the specific compression models w.r.t. specific tolerance extend of decoding complexities, the technique of variable-complexity is required.
For neural image compression, some recent works build variable-rate image compression frameworks\cite{Cai2019VBR, choi2019variable,9578818,song2021variable}, \textit{i.e.}, a unified model supporting compression at different bitrates. In the same way, a variable-complexity compression model can be devised.
We note that both the issues of variable-rate and variable-complexity compression are not typical in traditional image codecs, but they affect the practical deployment of neural image codecs significantly. Here, by adopting conditional transform modules, we build a unified model supporting variable-complexity compression. 

Our proposed variable-complexity compression method is a generalized work based on the widely-used spatial autoregressive context model~\cite{minnen2018joint}. For a source image $\boldsymbol{x} = \left [ x_{i} \right ]_{i=1:N}$, our model takes a complexity level $l\in [0, 1]$ as the side information to instruct the transformation and the probability estimation in latent represantation, which is related to the desired decoding complexity for user consumption. Additionally, the discrete entropy of $\boldsymbol {\hat z}$ is estimated conditioned on $l$, due to the difference of the distribution of $\boldsymbol {\hat z}$ under different time constraints. On the decoder side, the spatial binary mask $\boldsymbol M$ is decoded from $\boldsymbol {\hat z}$ to localize which spatial position should abandon the use of mask convolution.
Consequently, $\boldsymbol {\hat y}$ could be decoded utilizing the probability distribution calculated by both hyperprior and spatial-partially autoregression. Besides, since the rounding operation is not differentiable, uniform noise~\cite{balle2017end} is applied during training as an alternative.

\begin{figure}[t]
  \centering
  \includegraphics[width=0.94\linewidth]{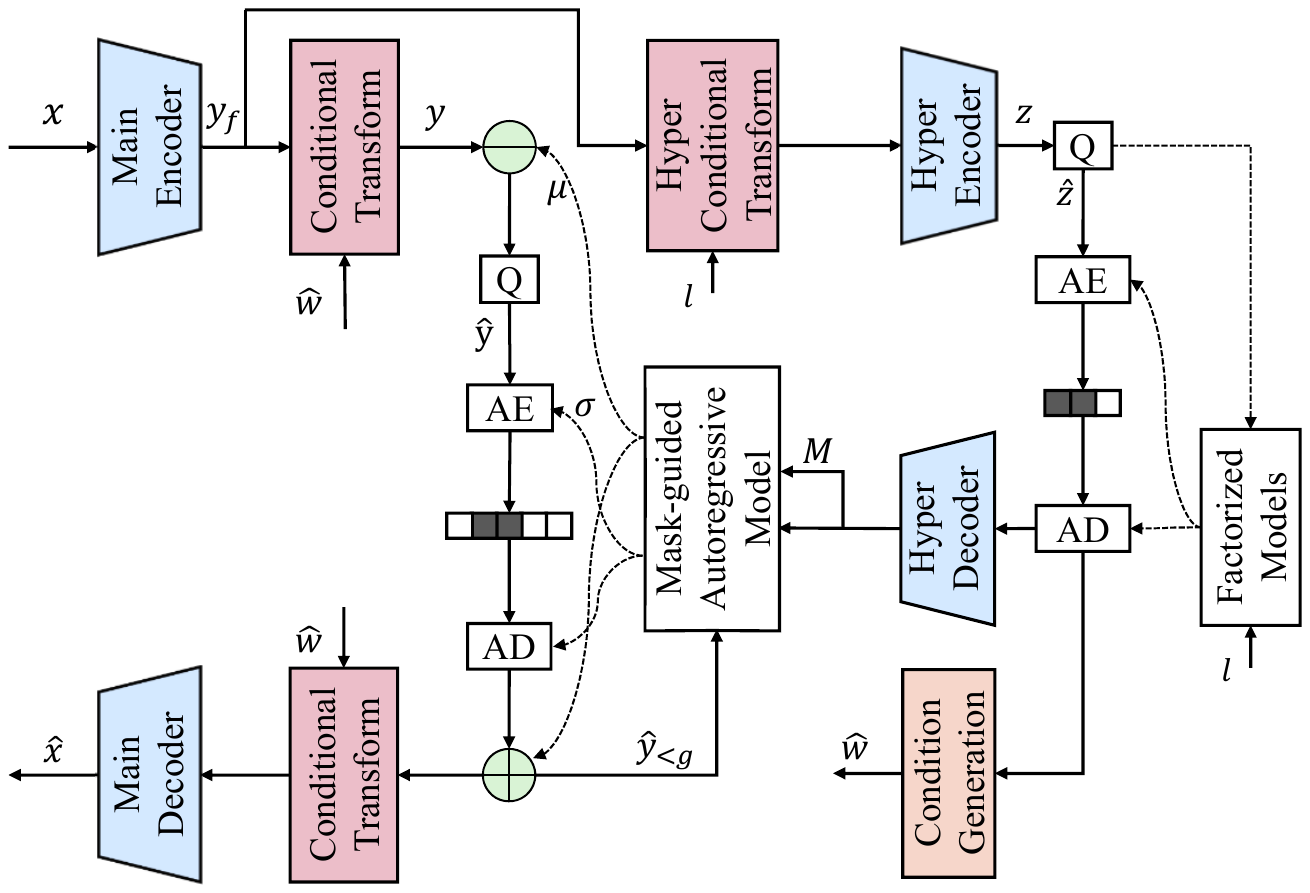}
  \caption{High-level structure of our variable-complexity model. According to an input complexity level $l$, our model generates a mask $M$ to indicate which positions in latents should be predicted by both autoregressive and hyperprior models.}
  \label{fig:VT_HighLevel_structure}
\end{figure}

Fig.~\ref{fig:VT_HighLevel_structure} shows the high-level structure of our proposed variable-complexity model. Section~\ref{sec:two-stage Training Strategy} and Section~\ref{sec: NetworkStructure} introduce the two-stage training procedure and the detailed implementation of our network architecture, respectively.

\begin{figure*}[ht]
  \centering
  \includegraphics[width=1.0\linewidth]{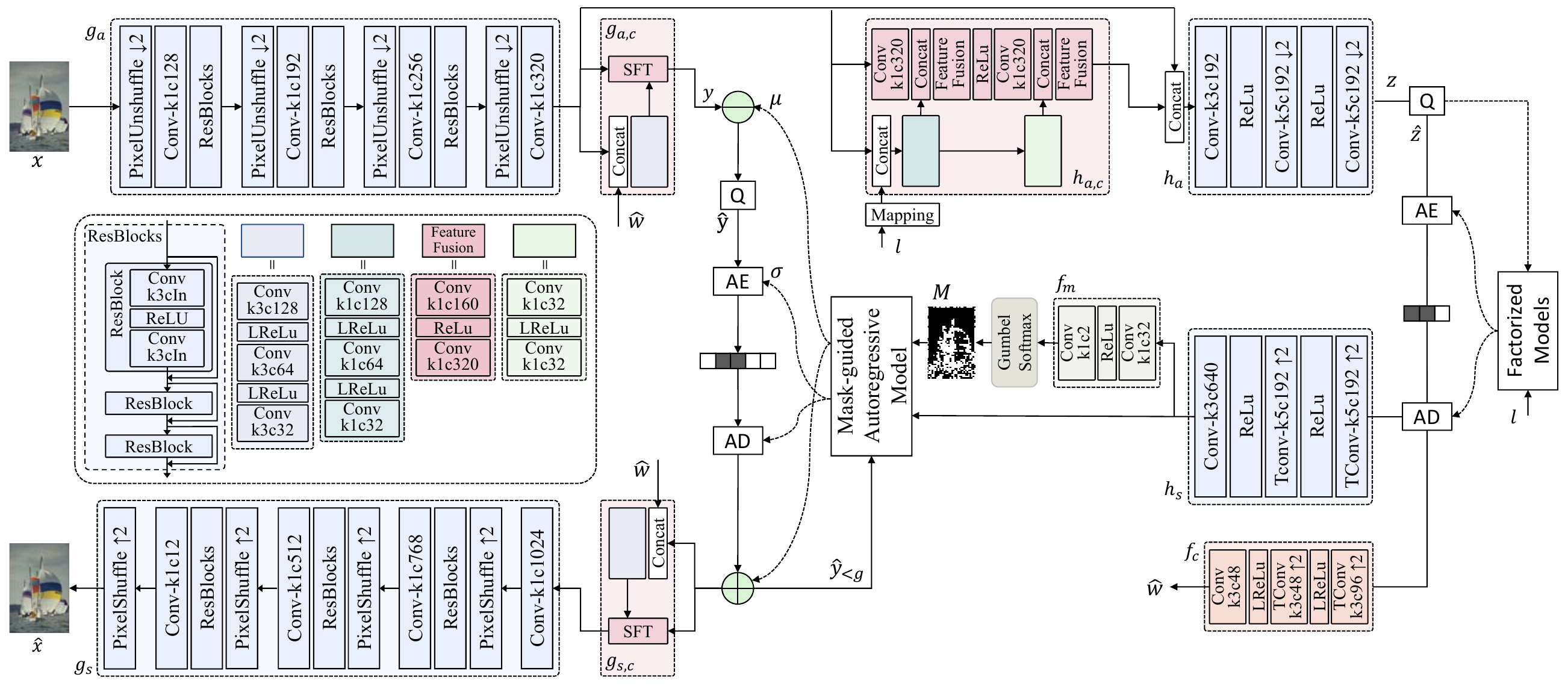}
  \caption{The network architecture of our proposed variable-complexity model. All the convolution settings are shown in the form of kernel size $k$ and output channel number $c$. Besides, $g_{a}$, $h_{a}$, $g_{s}$, $h_{s}$ denote the encoder, the hyper encoder, the decoder, the hyper decoder.
  $g_{a,c}$, $g_{s,c}$ and $h_{a,c}$ are all the corresponding conditional transform modules. AE and AD denote arithmetic encoding and arithmetic decoding and $Q$ denotes quantization. LReLU represents LeakyReLU.} 
  \label{fig:VTFramework} 
\end{figure*} 

\subsection{A Two-stage Training Strategy}
\label{sec:two-stage Training Strategy}
Motivated by the great success of masked image modeling, such as BeiT~\cite{bao2021beit}, MAE~\cite{he2022masked}, and SimMIM~\cite{xie2022simmim}, we incorporate the idea of pretraining a model with random masks and design a two-stage training strategy. 
Using this strategy, a pre-trained model that supports arbitrary spatial mask input will first be optimized. Then, to achieve better RDC tradeoffs at different decoding complexity levels, specific masks are learned.

More concretely, in the first stage, the spatial binary mask $M$ in Fig.~\ref{fig:VT_HighLevel_structure} is randomly sampled instead of being decoded from $\boldsymbol {\hat z}$. The ratio of 
points with a value of 1 in $M$ is from $0\% \sim 100\%$. 
% fine-grained spatial constraint
Nevertheless, we apply fine-grained spatial control inspired by the work~\cite{song2021variable}, and the training loss in the first stage is given by
\begin{equation}
L = -log P\left ( \boldsymbol{\hat{y}|\boldsymbol{M} } \right ) + \sum_{i}^{N}\frac{ \lambda_{D_{i}} }{N}\left ( x_{i} - x'_{i} \right )^2,
\label{eq:1stTrainingLoss}
\end{equation}
where $\lambda_{D_{i}}$ is the pixel-wise Lagrange multiplier and is varied by the spatial mask $M$. We upscale $M$ by 16 times and get $M'$, which has the same spatial resolution as the original image. Then the definition of $\lambda_{D_{i}}$ can be formulated as
\begin{equation}
\lambda_{D_{i}}=\left\{\begin{matrix}
 \lambda_{D} & ,M'_{i} = 0 \\
 \lambda_{D} \cdot s & ,M'_{i} = 1.
\end{matrix}\right.
\label{eq:lambda_D_i}
\end{equation}
Note that $s$ is a pre-defined value and is set to 0.9.
With the aid of the optimization goal Eq.~\ref{eq:1stTrainingLoss}, our proposed framework distinguishes the usage of mask convolution by measuring each pixel distortion. Through this training process, we can obtain a pre-trained compression model that is robust to changes in spatial masks.

In the second stage, we fix the main encoder and main decoder and optimize the mask generation. A better RDC tradeoff is pursued by learning the content-adaptive spatial mask $M$ and jointly optimizing the rate, distortion, and complexity constraints as
\begin{equation}
L = -log P\left ( \boldsymbol{\hat{y}|\boldsymbol{M} } \right ) + \sum_{i}^{N}\frac{ \lambda_{D_{i}} }{N}\left ( x_{i} - x'_{i} \right )^2+\lambda_{C}\cdot C_{e},
\label{eq:2ndTrainingLoss}
\end{equation}
where $\lambda_{C} = F(l)$ and $F$ is pre-defined function which is detailed description in appendix. 
In contrast to the loss function in the first stage, the constraint of the complexity indicator $C_e$ forces the model to determine where to use mask convolution based on the local dependencies in latents at a certain decoding complexity level. 
After the whole training process, a unified variable-complexity model is obtained.

\subsection{Network Architecture and Pipeline}
\label{sec: NetworkStructure}
Fig.~\ref{fig:VTFramework} demonstrates our proposed architecture.
We insert three conditional transform modules $g_{a,c}$, $g_{s,c}$, $h_{a,c}$ into the analysis transform, synthesis transform, and hyper encoder to adapt to the variation of decoding complexity levels.

We first transform the origin image $\boldsymbol x$ into $\boldsymbol{y_f}$ with the main encoder $g_a$. Then the hyper latent $\boldsymbol z$ is generated by a given complexity level $l$, which is given by 
\begin{equation}
\boldsymbol z = h_{a}(\boldsymbol{y_f}, \boldsymbol{\psi}) \text{, where } \boldsymbol{\psi} = h_{a,c}(\boldsymbol{y_f}, l).
\end{equation}
Since $l$ is a scalar, before being fed into the network, we broadcast it to a uniform spatial map. 
Due to the different complexities corresponding to the different amounts of hyper information to be transmitted, the distribution of quantized latent representation ${\boldsymbol {\hat z}}$ changes with the complexity level $l$. We exploit several factorized models to cover the variety range and connect $l$ with the index of the probability model $i$:
\begin{equation}
p_{{\boldsymbol {\hat z}}} ({\boldsymbol {\hat z}}, l) = p_{{\boldsymbol {\hat z}}}^{i} ({\boldsymbol {\hat z}})\text{, where } i = I(l), 
\label{eq:segmentFatorized}
\end{equation}
where $I(l)$ is a pre-defined segmented function and will be described in detail in the appendix.
After ${\boldsymbol {\hat z}}$ are transmitted to the decoder side, besides generating features for drawing the entropy parameters $(\mu,\delta)$, the spatial mask $M$ is also decoded from $\boldsymbol {\hat z}$ to guide the usage of the spatial autoregressive model. Moreover, an upsampling network $f_c$ is applied to $\boldsymbol {\hat z}$ to output the transform condition feature $\boldsymbol {\hat w}$, which maintains the information of decoding complexity.

After that, the modulated and quantized latent representation $\boldsymbol {\hat y}$ according to the complexity condition $\boldsymbol {\hat z}$ will be losslessly transmitted. 
For the conditional main transform, we apply the SFT module \cite{song2021variable}, which performs an element-wise affine transform to the input feature based on a prior condition. Finally, we obtain the reconstructed image $\boldsymbol {\hat x}$ with the conditional synthesis tranform $g_{s,c}$ and the synthesis transform $g_{s}$.  

\section{Experiments}
\label{sec:Experiments}
\subsection{Experimental Settings}
\label{sec:ES}
\begin{figure*}[t]
	\centering
	\subfloat[]{
	    \includegraphics[width=3.5in]{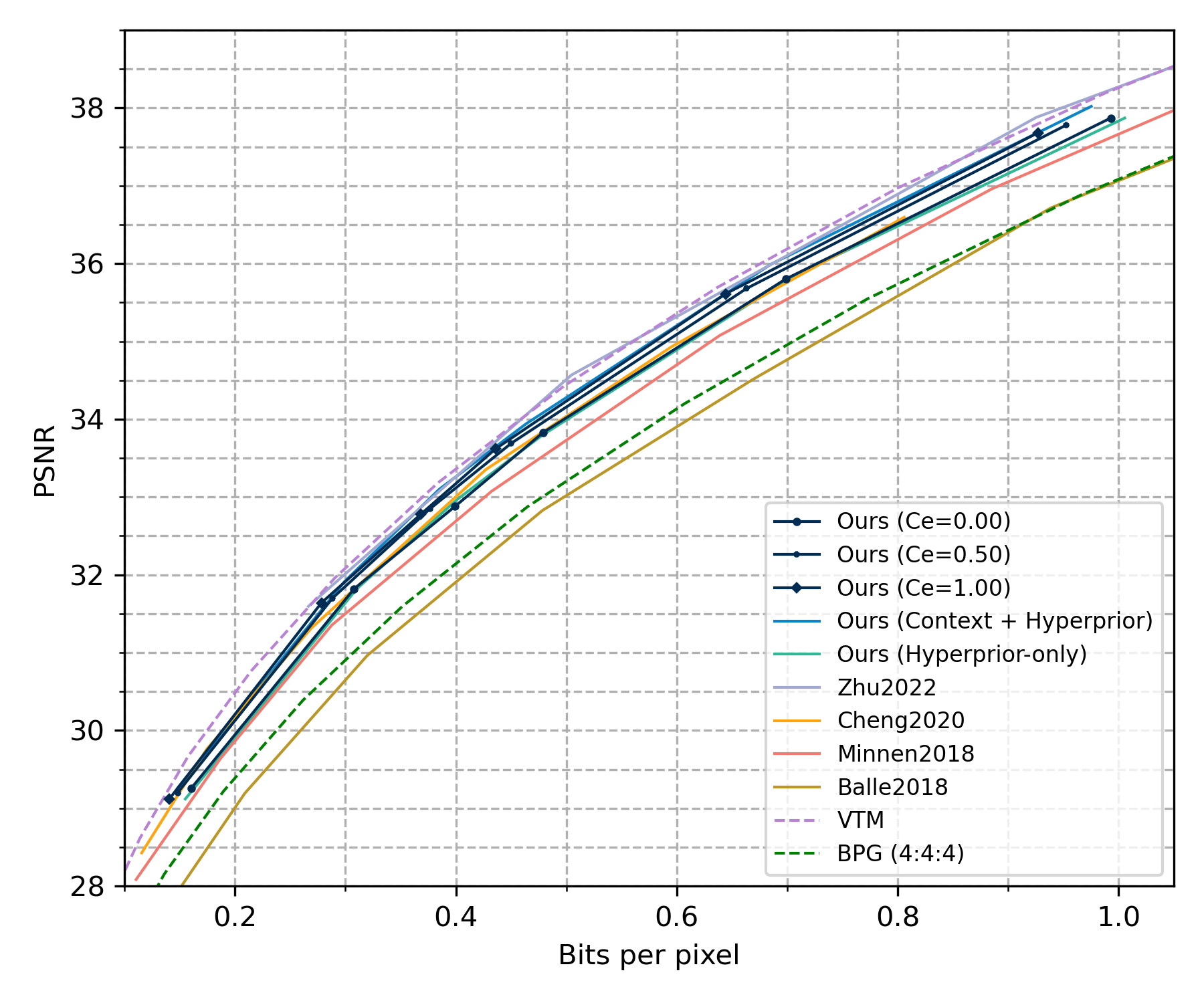}%
		\label{fig:RD_curve_vs_other_codecs}}
	\hfil
		\subfloat[]{
	    \includegraphics[width=3.5in]{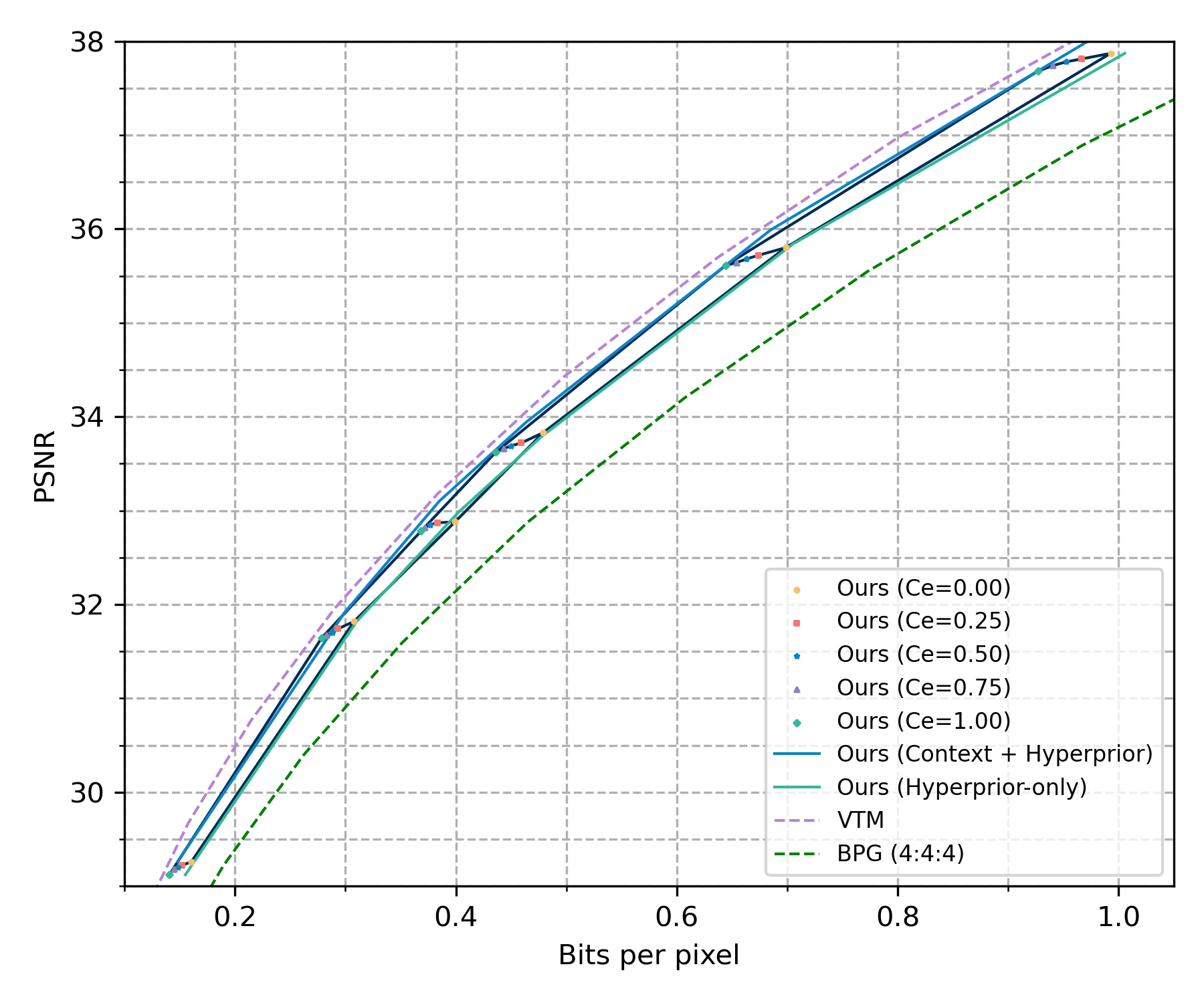}%
		\label{fig:RD_curve_vs_different_complexity}}
	\caption{Comparison of compression efficiency on Kodak. (a) compares the RD performance at three complexity levels of our model with other baseline models and (b) compares the relative performance of different decoding complexity levels in our model. The hyperprior-only and context + hyperprior models are the baseline models with the lowest and highest complexity, respectively.}
	\label{ContextModel}
\end{figure*}
\myparagraph{Training.} 
We choose the COCO 2017 dataset~\cite{lin2014microsoft} as our training set and each iteration randomly selects eight 256 $\times$ 256 crops from the training images. Our distortion $D$ in Eq. \ref{eq:RDC_goal} is MSE. The models are optimized using Adam optimizer~\cite{kingma2014adam} with six different $\lambda_{D} \in \left \{ 128, 512, 768, 1024, 2048, 4096\right \}$. We first train a hyperprior-only model at $\lambda_{D} = 16384$ from scratch for 1M steps with learning rate $5e-5$ as the starting point. Then our proposed two-stage training strategy is applied. First, we load the weights from the pre-trained high-rate hyperprior-only model with random masks with 1.5M iterations. Then we fix the main encoder and main decoder and finetune our model for 0.6M iterations. The learning rate is initially set to $5e-5$ and then decreased to $1e-5$ in the first and second stages at 1M and 0.3M iterations, respectively. All the experiments are conducted on an RTX 3080Ti GPU with Pytorch platform.

\myparagraph{Testing.}
The models are evaluated on Kodak dataset \cite{franzen1999kodak} and Tecnick dataset \cite{asuni2014testimages}. Kodak dataset contains 24 512x768 RGB images and Tencnick dataset contains 100 1200x1200 RGB images. The decoding complexity is evaluated on an RTX 1080Ti GPU. 
To compare with traditional codecs, we convert the images to YUV color space and use BPG~\cite{bellard2015bpg} and VTM-18.2~\cite{ohm2018versatile} to code the images in YUV444 mode, and then calculate PSNR in RGB.
We also compare our method with four neural image compression methods, Zhu2022~\cite{zhu2021transformer}, Cheng2020~\cite{cheng2020learned}, Minnen2018~\cite{minnen2018joint}, Balle2018~\cite{balle2018variational}.
% We also provide the rate-distortion curves as well as BD-rate \cite{bjontegaard2001calculation} savings against the-state-of-art benchmarks \cite{minnen2018joint,chen2020learned}.
 
\subsection{Quantitative results}
\label{sec:VC}
In this section, we conduct experiments to evaluate and analyze the performance of our variable-complexity model. To make a fair comparison, we additionally optimized two baseline models with the lowest decoding complexity (hyperprior-only) and the highest decoding complexity (context + hyperprior), respectively. Both of them implement the same network architectures of $g_{a}$, $g_{s}$, $h_{a}$, $h_{s}$ and $g_{ep}$ of ours proposed model, as shown in Fig.~\ref{fig:VTFramework}. 
The remaining parts of Ours (Hyperprior-only) and Ours (Context + Hyperprior) are the same as ``Mean \& Scale Hyperprior'' and ``Context + Hyperprior''~\cite{minnen2018joint}, respectively. Both are trained based on the pre-trained high bitrate hyperprior-only model for 1M iterations with a learning rate of $5e-5$ and 0.5M iterations with a learning rate of $1e-5$, which is the same as the first stage training schedule of our variable-complexity mentioned in Section~\ref{sec:ES}.

\myparagraph{Rate-distortion curves.}
In the test phase, decoding time is able to be finely adjusted by varying the complexity level $l \in [0, 1]$ within a single model. However, to easily and clearly conduct the performance comparison, we manually select several representative entropy decoding complexity degrees %$C_e \in \left \{ 0, 0.25, 0.5, 0.75, 1 \right \}$ 
(noted as Ours ($C_e$=0), Ours ($C_e$=0.25) and so on) and evaluate the rate-distortion performance. 

As shown in Fig~\ref{fig:RD_curve_vs_other_codecs}, the RD performance of our model improves as the decoding complexity increases from low to high.
Compared to the two baseline models, our proposed variable-complexity model has comparable performances, which only a slight performance drop at the highest-complexity bound ($C_e$=1.0). Since it realizes decoding complexity control, we think the quality loss is acceptable.
In addition to providing the RD curves at the two complexity extremes, we also present the performance of the model with a $0.5$ complexity indicator. Ours ($C_e$=1.0) and Ours ($C_e$=0.5) achieve similar performance, demonstrating that the performance gain of the spatial autoregressive model mainly comes from the effective context prediction of part of the content.

Fig.~\ref{fig:RD_curve_vs_different_complexity} shows the RD performance between different complexity degrees of our model. We can see that even if only 25\% of the latent points require the usage of mask convolution ($C_e$=0.25), it can already bring a performance improvement of nearly half the adjustable range. Moreover, as decoding complexity increases, the RD performance of our model improves increasingly, up to the pre-defined upper bound of complexity.

\paragraph{Decoding complexity}
\begin{table*}[!t]
\begin{center}
\caption{Total decoding time averaged on Kodak and Tecnick (ms). All the models have a $H \times W \times 320$ feature map to transmit. The concrete value of $H$ and $W$ depends on test dataset which is denoted below.}
\label{tab:DecodingTime}
\begin{tabular}{@{}cccccccc@{}}
\toprule
\multicolumn{1}{c}{\multirow{2}{*}{Model}} & Ours (Hyperprior-only) & \multicolumn{5}{c}{Ours} & Ours (Context + Hyperprior) \\
\multicolumn{1}{c}{}                       
& (Parallel)             
& \begin{tabular}[c]{@{}c@{}}$C_{e}$=0.0\\ (\textless{}0.01)\end{tabular}          
& \begin{tabular}[c]{@{}c@{}}$C_{e}$=0.25\\ ($\pm$ 0.01)\end{tabular}
& \begin{tabular}[c]{@{}c@{}}$C_{e}$=0.5\\ ($\pm$ 0.01)\end{tabular}
& \begin{tabular}[c]{@{}c@{}}$C_{e}$=0.75\\ ($\pm$ 0.01)\end{tabular}
& \begin{tabular}[c]{@{}c@{}}$C_{e}$=1.0\\ (\textgreater{}0.99)\end{tabular}
& (Serial)
\\ \midrule
Kodak(768$\times 512$) & 114.00 & 176.00 & 1641.00 & 3025.33 & 4537.33 & 7283.33 & 7227.50 \\
Tecnick(1200$\times 1200$) & 410.33 & 658.17 & 6715.50 & 11779.25 & 16715.83 & 21684.33 & 21568.67 \\
\bottomrule
\end{tabular}
\end{center}
\end{table*}

We provide the average decoding time of the two baselines (hyperprior-only and context + hyperprior) and different decoding complexity degrees of our variable-complexity model ($C_e$=0.00, 0.25, 0.50, 0.75, 1.00) on Kodak and Tecnick datasets.
As shown in Table~\ref{tab:DecodingTime}, the decoding time of Ours ($C_e$=0.0) is about 1.5 times that of the hyperprior-only model, mainly due to the additional computation cost of mask generation and a few autoregressive process calculations.
Meanwhile, Ours ($C_e$=1.0) has a similar decoding time to the context + hyperprior model, which indicates that the autoregressive operation dominates the decoding complexity. In our model, as the selected complexity increases, the decoding time significantly increases, and these two are linearly correlated. This also verifies the rationality of quantifying decoding complexity using autoregressive times.

\myparagraph{BD-rate versus decoding time.}
To further evaluate the rate-distortion-complexity performance, Fig.~\ref{fig:BD_rate_vs_DecodingTime} shows the relationship between BD-rate and decoding time of various compression methods. We provide the BD-rate vs decoding time curve of our proposed variable-complexity model and the corresponding points of several state-of-the-art learning-based methods. We calculate the BD-rate savings of our model with different complexity levels against VTM 18.2. Our proposed variable-complexity model can finely change the decoding time from 176ms to 7283.33ms on Kodak dataset, where the adjustable BD-rate range is about 6\%. 

\myparagraph{Model complexity.}
Table~\ref{Tab:ModelComplexity} demonstrates the model parameters of our baseline models (hyperprior-only and context + hyperprior) and the proposed model (Ours). 
The parameter difference between the context + hyperprior and the hyperprior-only model mainly comes from mask convolution. Compared with the context + hyperprior model, our variable-complexity model additionally inserts the conditional transform modules ($g_{a,c}$, $g_{s,c}$, $h_{a,c}$), the mask generation module ($f_{m}$), and the condition generation module ($f_{c}$) as shown in Fig.~\ref{fig:VTFramework}. These cause an increase in 4.1M parameters.
\begin{table}[t]
\begin{center}
\caption{Comparison of model complexity.}
\begin{tabular}{cccc}
\toprule
Model  & \begin{tabular}[c]{@{}c@{}}Ours \\ (Hyperprior-only)\end{tabular}  
& \begin{tabular}[c]{@{}c@{}}Ours \\ (Context + Hyperprior)\end{tabular} 
& Ours 
\\ \midrule
Params & 22.26M & 27.39M & 31.50M
\\ \bottomrule
\label{Tab:ModelComplexity}
\end{tabular}
\end{center}
\end{table}

\subsection{Analysis}
\myparagraph{Spatial mask.}
\begin{figure*}[ht]
  \centering
  \includegraphics[width=1.0\linewidth]{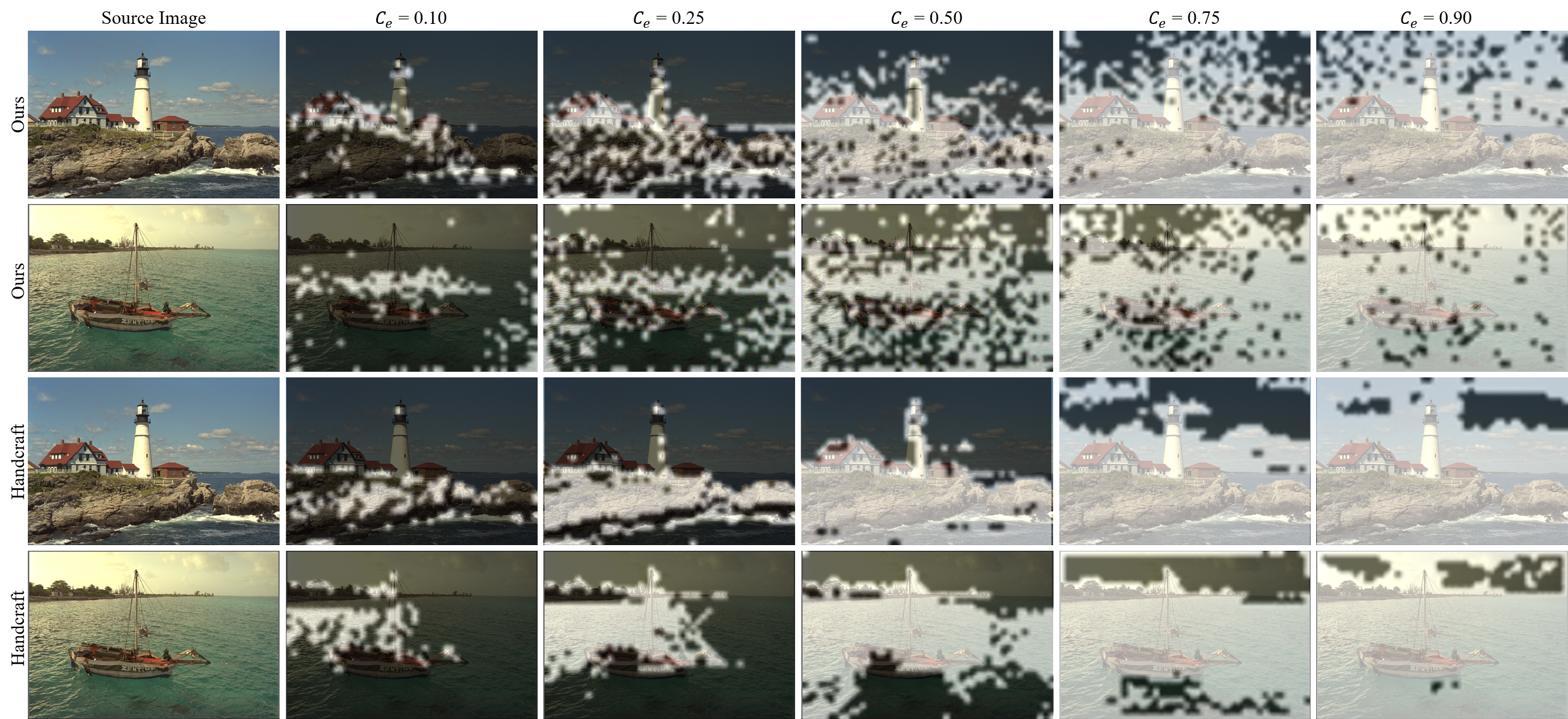}
  \caption{Visualization of the spatial masks $M$ at the different decoding complexity degrees $C_e$ of entropy model. The bright points will be predicted by both the context model and the hyperprior, while the dark ones are only predicted by the hyperprior.}
  \label{fig:visualmask}
\end{figure*}
To illustrate how the entropy model chooses available context information, we visualize the spatial masks $M$ at different entropy decoding complexity indicators $C_e$ within a single variable-complexity model, which measures the proportion of latent positions chosen using mask convolution. The first column shows the source images and the following columns correspond to different $C_e$ values. In Fig.~\ref{fig:visualmask}, the bright regions are chosen and the dark ones are masked. 
We can see the local spatial context of object edges (including the image edge) is given priority from the second column ($C_e$=0.1). As the complexity increases, the local spatial redundancy of the regions with complex texture is further extracted, while the regions with regular textures are considered later (\textit{e.g.} the blue sky excluding white cloud in the first two rows at $C_{e}=0.5$). This is consistent with the characteristic that the context model can preserve more details. 

\myparagraph{Latent entropy.} 
\begin{figure}[t]
    \centering
    \includegraphics[width=1.0\linewidth]{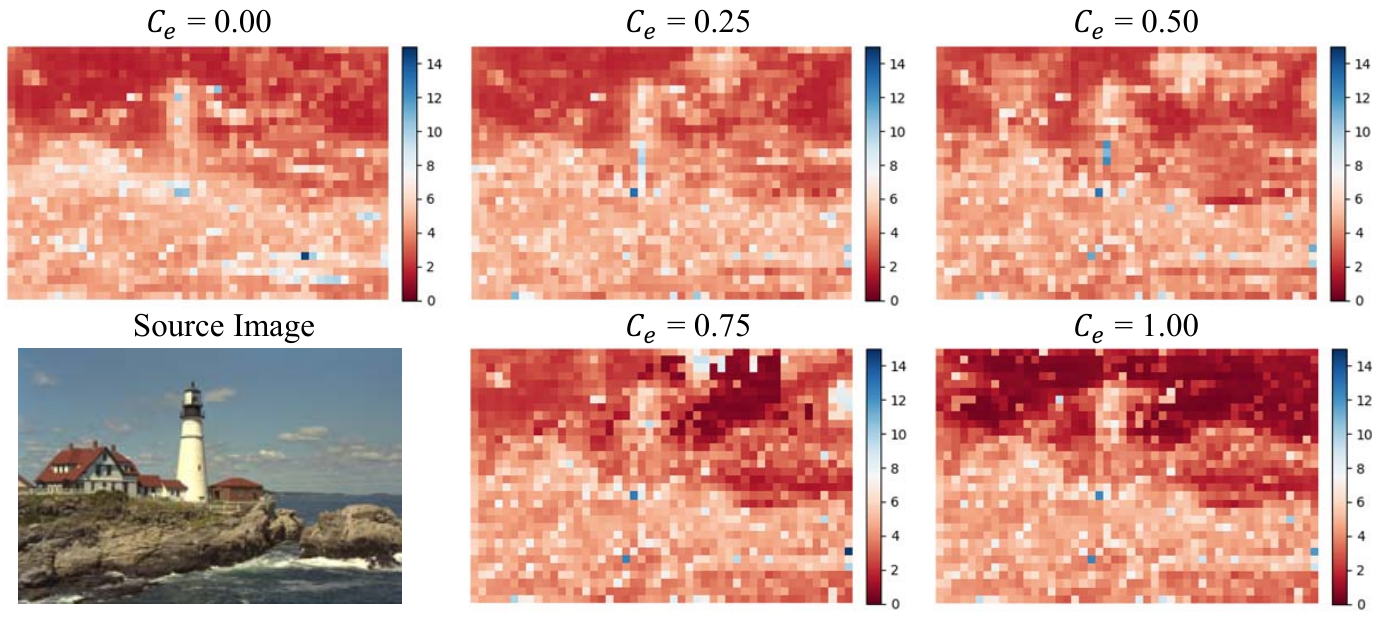}
    \caption{Visualization of latent entropy for the channel with the highest entropy at different decoding complexity levels in our model.}
    \label{fig:Visual_RequiredBits}
\end{figure}
Fig.~\ref{fig:Visual_RequiredBits} visualizes the variation of latent entropy for the channel with the highest entropy according to different decoding complexity levels in our model. 
With the increasing use of mask convolution in the latent representation, there is a noticeable reduction in bits cost for the lighthouse and house, rocks, blue sky, and clouds, respectively.

\myparagraph{Learned mask versus handcraft mask.}
\begin{figure}[t]
    \centering
    \includegraphics[width=1.0\linewidth]{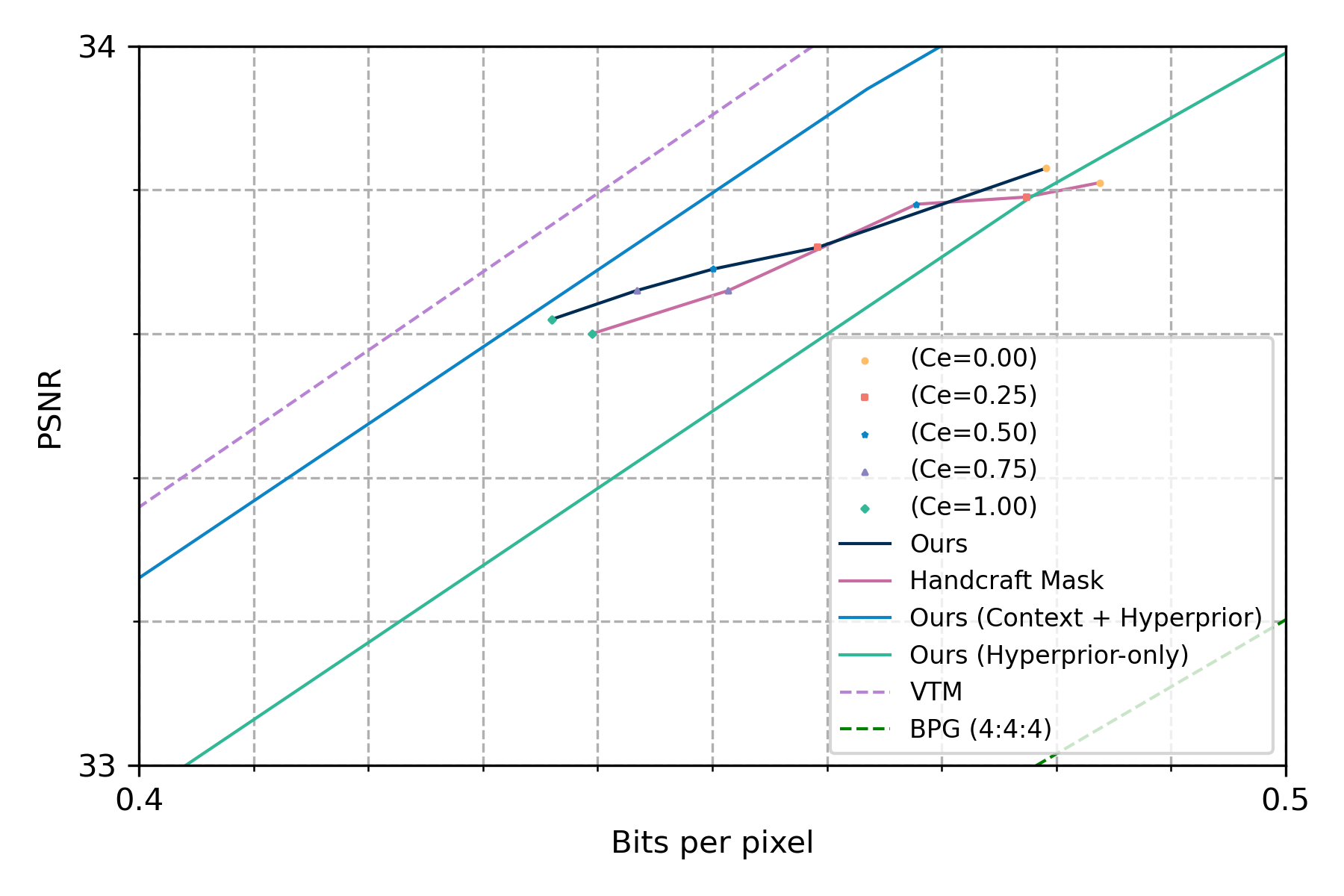}
    \caption{Comparison between learned and handcraft mask.}
    \label{fig:learn_vs_handcraft_mask}
\end{figure}
We compare the effectiveness of two mask generation forms, learned or handcrafted. In fact, after the first training stage, our model already has the potential to arbitrarily adjust decoding complexity. 
If introducing a manually designed mask generation method based on our pre-trained model can achieve good RD performance, then the necessity of the second stage training will vanish.
To explore this issue, we consider that latent points not well predicted by the hyperprior model alone may require the help of mask convolution, and employ a corresponding handcrafted mask generation algorithm.
At the encoder side, before transmitting the latent representation $\boldsymbol{\hat{y}}$, the spatial-wise entropy of it is first calculated only by the hyperprior model. Then, we use the probability parameters from the hyperprior model to code a ratio of $1 - C_e$ latent points with lower entropy. The remaining higher entropy latent representation will be transmitted by both the context and hyperprior model. Unlike our learned mask generation, the handcrafted mask requires a few overheads for transmission.

Fig~\ref{fig:learn_vs_handcraft_mask} compares the RD performance of the two mask generation schemes. It can be seen that our model performs obviously better than handcrafted masks at each level of decoding complexity degrees. Moreover, under the guidance of handcrafted masks, an additional 50\% entropy decoding complexity cost only results in a bitrate savings of about 3.2\%, which is even much worse than our model with learned masks at 25\% complexity degree ($C_e$=0.25). The bottom two rows in Fig~\ref{fig:visualmask} show the spatial masks under the handcrafted scheme at different complexity degrees. In comparison to our proposed model (the top two rows), the mask appears much flatter and does not capture the differences between local spatial textures. This is also a problem with the hyperprior model. Therefore, learning mask generation is necessary and can bring significant performance gains.

\subsection{Ablation Study}
\begin{figure}[h]
    \centering
    \includegraphics[width=1.0\linewidth]{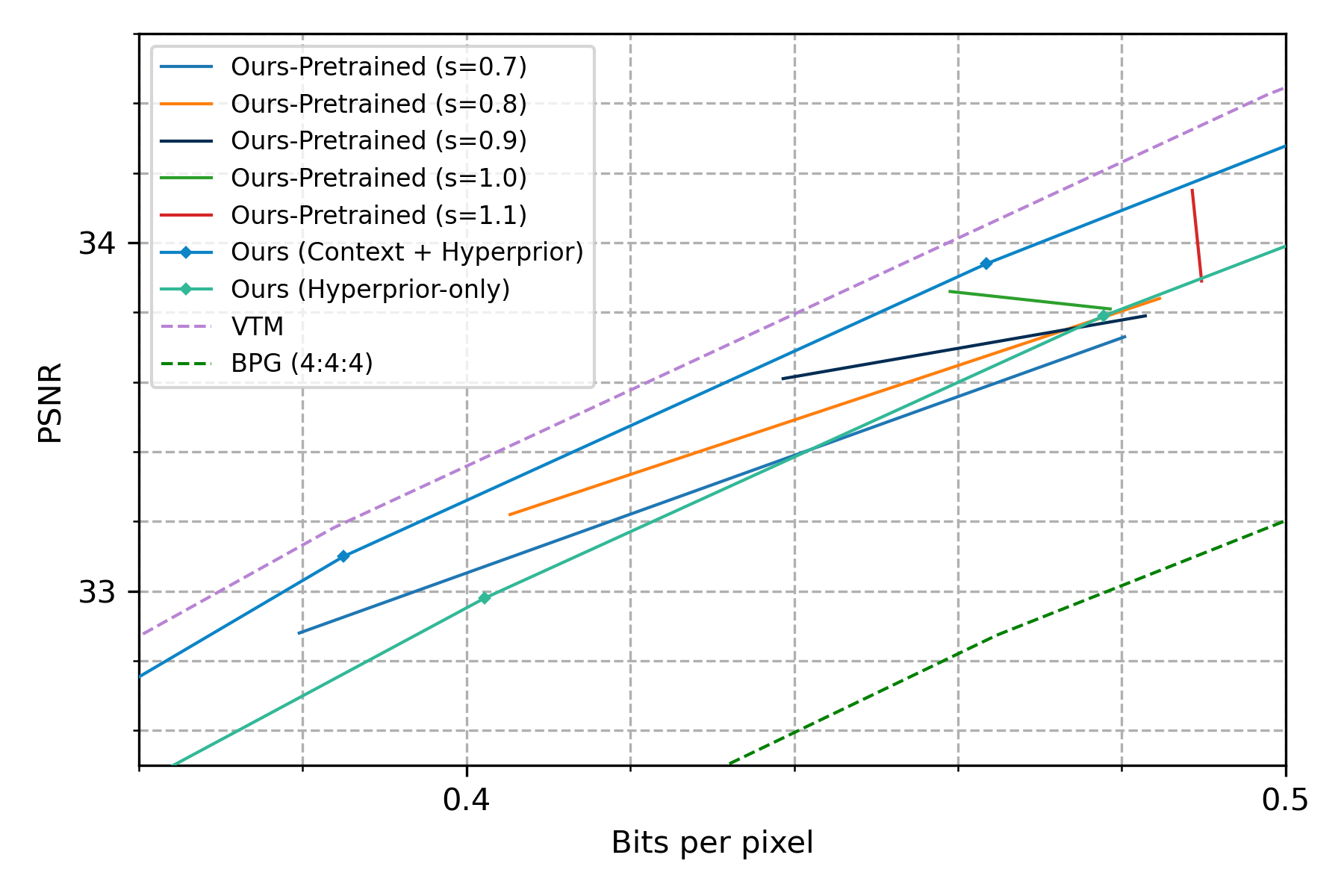}
    \caption{Ablation study.}
    \label{fig:Ablation_mcscale}
\end{figure}
Fig.~\ref{fig:Ablation_mcscale} depicts the effectiveness of the scale $s$ in the training loss (Eq.~\ref{eq:lambda_D_i}). In the range of $s$ from 1.1 to 0.7, the range of bitrate that one model can cover increases, but the performance degradation also becomes increasingly apparent. Therefore, the scale $s$ can be seen as a trade-off between RD performance and the adjustable bitrate range. In this work, we make a balanced choice and set $s=0.9$. When users have other requirements (\textit{e.g.} better RD performance), they can achieve it by adjusting $s$.

\section{Conclusions} 
In this paper, we systematically investigate the rate-distortion-complexity (RDC) optimization in neural image compression. For the first time, we enable precious control of the rate-distortion-complexity trade-off by quantifying and incorporating the decoding complexity into optimization. Moreover, a variable-complexity neural codec is proposed to support fine-grained complexity adjustment, which adaptively controls the spatial dependencies modeling in the context model. Our comprehensive experimental results demonstrate the feasibility and flexibility of RDC optimization for neural image codecs.

\bibliographystyle{IEEEtran}
\bibliography{egbib}

\clearpage
\appendix
\subsection{Network architecture details}
\begin{figure}[ht]
  \centering
  \includegraphics[width=0.95\linewidth]{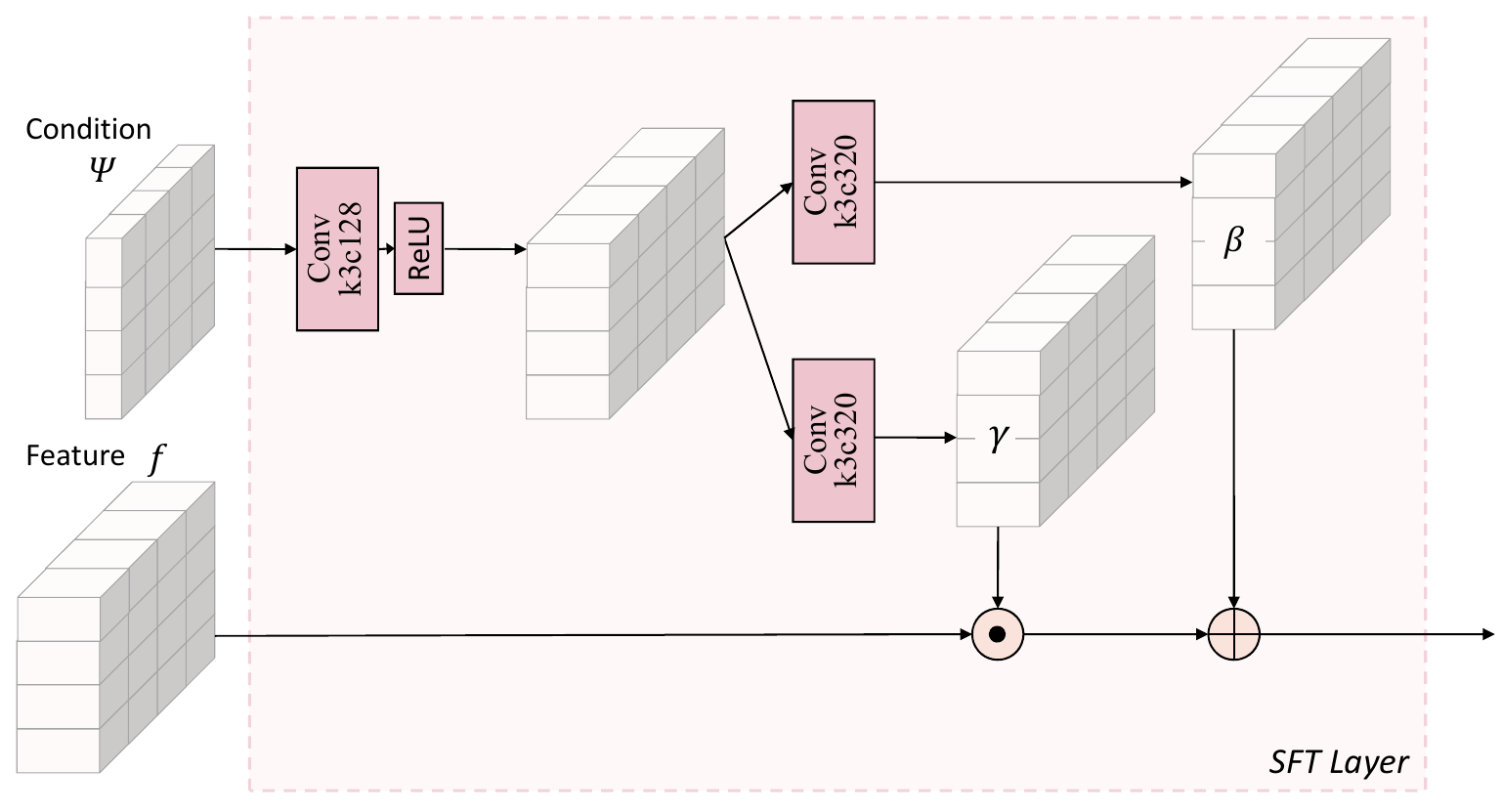}
  \caption{Illustration of Spatial Feature Transform (SFT) layer~\cite{song2021variable}. SFT layer performs an element-wise affine transformation on the input feature $f$.} 
  \label{fig:SFT}
\end{figure}

\myparagraph{Spatial Feature Transform.}
We conduct our condition main transform through Spatial Feature Transform (SFT) module in the work~\cite{song2021variable}, which performs an element-wise affine transformation on the input feature conditioned on the condition feature. Fig.~\ref{fig:SFT} illustrates the concrete structure of the SFT layer.

The SFT will generate a set of element-wise affine parameters $\gamma$ and $\beta$ from an external condition $\Psi$ for the feature $f$. The SFT layer transformation is given by:
\begin{equation}
SFT(f,\Psi ) = \gamma \odot f+\beta.
\label{eq:SFT}
\end{equation}
where $\odot$ represents element-wise multiplication.

\myparagraph{Segmented factorized models.}
In Eq.~\ref{eq:segmentFatorized}, we mention that we apply several factorized models and choose which one to use according to the complexity level $l$. Concretely, 12 factorized models are exploited to handle 12 segments from the full-range complexity level separately:
\begin{equation}
\begin{aligned} 
\relax [0,1]= [0,0.01) \cup [0.01,0.1) \cup [0.1,0.2) \cup [0.2,0.3) \\ \cup [0.3,0.4) \cup [0.4,0.5) \cup [0.5,0.6) \cup [0.6,0.7) \\ \cup [0.7,0.8) \cup [0.8,0.9) \cup [0.9,0.99) \cup [0.99,1].
\end{aligned}
\end{equation}

\subsection{Training loss}
In Eq.~\ref{eq:2ndTrainingLoss}, the definition of function $F(l)$ is:
\begin{equation}
    F(l) = \frac{1}{256}\left ( a + b \cdot l + c \cdot l^2 + d \cdot l^3 + e \cdot l^4 \right ),
\end{equation} 
where the coefficient values regarding different ${\lambda}_{D}$ is shown in the following table:
\begin{table}[h]
\begin{center}
\caption{Coefficient values regarding different ${\lambda}_{D}$.}
\begin{tabular}{cccccc}
\toprule
${\lambda}_{D}$ & a    & b     & c     & d     & e     \\ \midrule
192    & 15   & 68.4  & 126.6 & 100.4 & 27.3  \\
512    & 20   & 109   & 250.5 & 249   & 87.5  \\
768    & 24.9 & 145.8 & 343.3 & 336.4 & 113.5 \\
1024   & 25   & 140.8 & 336   & 344.6 & 124.2 \\
2048   & 35   & 181   & 400.2 & 388.4 & 133.9 \\
4096   & 39.8 & 242.2 & 625.9 & 684.2 & 260.5 \\ \bottomrule
\end{tabular}
\end{center}
\end{table}

\subsection{Traditional codec evaluation}
\label{sec:TraditionalCodecsSettings}
The evaluation scripts used to generate results for traditional codecs are as follows:

\myparagraph{VTM-18.2.} 
We build VTM-18.2\footnote{\url{https://vcgit.hhi.fraunhofer.de/jvet/VVCSoftware_VTM/-/tags/VTM-18.2}} software and use the script from CompressAI~\cite{begaint2020compressai} for dataset evaluation. Specifically, the following command is adopted to generate VTM-18.2 image compression evaluation results on Kodak and Tecnick datasets:
\begin{lstlisting}[language=python]
python -m compressai.utils.bench vtm [path to image folder] 
-c [path to VVCSoftware_VTM folder]/cfg/encoder_intra_vtm.cfg 
-b [path to VVCSoftware_VTM folder]/bin 
-q 16, 18, 20, 22, 24, 26, 28, 30, 32, 34, 36, 38, 40
\end{lstlisting}

\myparagraph{BPG.} 
We build BPG\footnote{\url{https://bellard.org/bpg/}} software from and use the following commands to encode and decode Kodak and Tecnick datasets:
\begin{lstlisting}
bpgenc -e x265 -q [0 to 51] -f 444 
-o [encoded heic file] [original png file] 
bpgdec -o [decoded png file] [encoded heic file]
\end{lstlisting}

% \begin{IEEEbiography}[{\includegraphics[width=1in,height=1.25in,clip,keepaspectratio]{photos/高白.jpg}}]{IEEE Publications Technology Team}
% In this paragraph you can place your educational, professional background and research and other interests.
% \end{IEEEbiography}
\end{document}